# More-than-Moore Microacoustics: A Scalable Fabrication Process for Suspended Lamb Wave Resonators

Marco Liffredo, *Member, IEEE*, Federico Peretti, Nan Xu, Silvan Stettler and Luis Guillermo Villanueva, *Member, IEEE*

***Abstract*—** Deep Ultraviolet (DUV) Photolithography is currently used to fabricate mass-scale integrated circuits (ICs). Its high throughput and resolution could benefit large-scale RF MEMS production for the telecommunication market. We present a process flow to fabricate suspended acoustic resonators using DUV Photolithography. This method allows for scalable production of resonators with critical dimensions of 250 nm and alignment accuracy of less than 100 nm. We show how photoresists and anti-reflective coatings integrate with the process, help with deposition quality and resolution, and how Ion Beam Etching allows for vertical sidewalls of the resonators. We measure resonance frequencies ($f_r$) up to 7.5 GHz and electromechanical couplings up to 8%, and we investigate the uniformity of this process by analyzing the deviation of $f_r$ over the wafer surface for four main resonance modes. We show that deviation of the S0 mode can be kept below 1%. These results indicate the suitability of this process for quick scale-up of Lamb wave resonator technology, bridging the gap from research to industry.

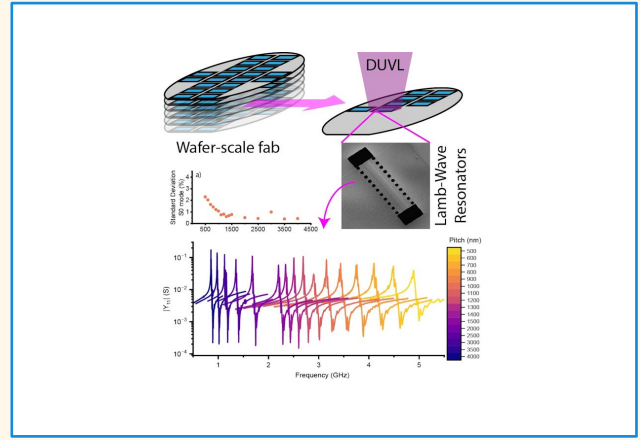

***Index Terms*—** Acoustic Devices, Aluminum Nitride, Microfabrication, Microelectromechanical devices, Photolithography

## I. INTRODUCTION

THE shift from 4 to 5G, and future developments of 6G opens the way to the exploitation of new frequency bands for wireless telecommunication. With the increase of connected wireless devices, the number of acoustic bandpass filters will keep growing. It is expected that the market size for RF-MEMS reaches $20B by 2028[1]. Many technologies developed for such market have matured and are widespread in the consumer electronics industry as elements for filters and oscillators[2]. Nowadays, filtering elements are primarily composed of two families of acoustic resonators: thin Film Bulk Acoustic Resonators [3], [4] (FBAR) made with AlN or Sc-doped AlN, and Surface Acoustic Wave (SAW) resonators made out of single crystal materials, such as LiNbO3 and LiTaO3. Single-crystal piezoelectric materials can achieve higher piezoelectric coupling than Sputter-deposited nitrides [5], but non-post-CMOS compatible technologies are required to transfer the piezoelectric film on a chip. Nitrides of the AlN family can be easily deposited with post-CMOS compatible Reactive Sputtering, and AlN's lower coupling can be compensated with rare-earth doping. For this reason, FBAR resonators have been traditionally used to operate at the higher frequency bands: the ease of controlling resonance frequency with the film thickness, [5], [6]. On the other hand, SAW devices define their resonance frequency with the pitch of interdigitated transducer electrodes (IDT). This allows the production of SAW resonators operating at different frequencies with one single lithography step. Lithography requirements for FBAR operating at high frequency are much more relaxed than for SAW since they do not require IDTs to define their wavelength, but this comes with the downside of requiring extra fabrication steps to change the resonance frequency and integrate multiple frequencies in the same chip, either with mass-loading or thickness trimming. With the push for increased coupling and bandwidth, new bulk wave resonators requiring IDT have been introduced in research and commercial applications. It is the case of Laterally eXcited Bulk Acoustic Resonators (XBAR)[7] where the frequency is thickness-defined or 2D mode resonators (2DMR)[8] where the lateral and the thickness vibration modes are hybridized in a 2D resonance mode to improve the piezoelectric coupling. In such devices, the maximum coupling is achieved with film thickness ratios over lateral wavelength (λx) around 0.4[9]. In general, to achieve high operation frequency with a lateral mode resonator, the main lithographic limitation is the precise patterning and aligning of the IDTs, whose width can reach dimensions less than 200 nm. There are different techniques to achieve such a small feature size, with the most used in academic environments being electron-beam (e-beam) lithography [11], [12]. The main

drawback of e-beam lithography is the serial writing of wafers, which restricts its use for large-scale production. Following the concept of "More-than-Moore"[10], technologies used for mass fabrication of integrated circuits (ICs) can be translated into MEMS production, benefitting from all the previous years of investments and R&D towards patterning of smaller and smaller features in a single chip. Deep Ultraviolet (DUV) Photolithography represents a possible solution for implementing pitch-defined MEMS devices operating at high resonance frequencies. It is an established technology with more than 30 years of development, and so costs are much lower compared to more modern technologies, such as Extreme Ultraviolet Photolithography[11]. A fully optimized DUV lithography cluster (combining coating, exposure, and development) allows for a throughput of hundreds of wafers per hour with a resolution of 150 nm and an overlay of less than 100 nm, matching this technology with the feature size requirements for small pitch IDT. Using such a mature technology helps reduce production costs because old tools used in producing old transistor nodes can be used for MEMS fabrication without requiring new equipment purchase. This paper demonstrates the first published process flow for producing Sc-doped AlN (AlScN) Lamb Wave resonators relying on DUV lithography to achieve high resolution and accurate alignment of metals and piezoelectric layers. To achieve large coupling, we use Sc-doped AlN with a Sc concentration of 40% at%, and we cover a frequency range from 500 MHz up to 7.5 GHz. In the paper, we analyze the reticle writing, resonator fabrication, and characterization, quantifying the performance uniformity across the wafer surface.

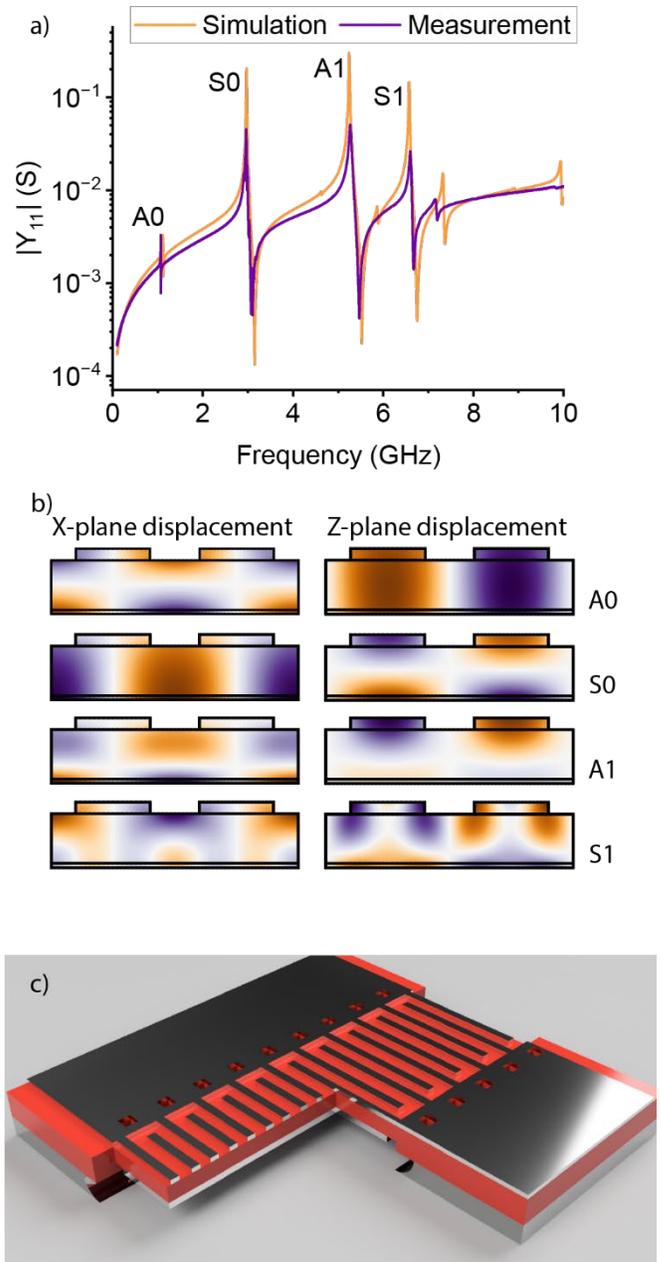

Fig 1. a) Comparison between simulation and measurements of a resonator with b) colormaps showing x-axis displacement and z-axis displacement for the four main resonance modes investigated in this work. c) a sketch of the resonator to show the section of the suspended structure and the electrode arrangement.

*Highlights*

- **We present a process flow to leverage the high throughput of Deep Ultraviolet photolithography to scale up production of Lamb wave resonators.**
- **We achieve overlay accuracy lower than 100 nm and frequency deviation below 1% over the wafer surface for the S0 resonance mode.**
- **The proposed process flow allows the combination of the flexibility in frequency selection of Lamb Wave Resonators, the high coupling of AlScN, and Wafer-scale production.**

## II. Device design

To show the full potential of DUV lithography, this paper focuses on Lamb wave resonators with resonance frequencies that can be easily set by lithography, and thus, we can target a large frequency range. The architecture of our Lamb wave resonator excites both symmetric (S0, S1) and asymmetric (A0, A1) Lamb modes, originating multiple resonance frequencies in the electric response (see Fig 1.a and b for the simulation of the mode shapes). A sketch of the devices is visible in Fig 1..c: a suspended $Al_{0.6}Sc_{0.4}N$ plate with a top IDT and a bottom floating electrode. The presence of the bottom metal is mainly to promote good growth of the piezoelectric layer (as explained in[12]). In addition, the floating bottom metal plate helps maximize the electric field's vertical component and improve transduction using both the $e_{31}$ and the $e_{33}$ coefficients. The downside of a bottom metal plate is the introduction of large parasitic capacitance. For this reason, the Pt layer is patterned to cover only the aperture of the resonators. The IDT pitch is swept from 500 nm up to 4.5 µm, resulting in a vast range of frequencies (from 750 MHz to 5 GHz for the S0 mode). We sweep over a large pitch to see the impact of a purely vertical electric field (where thickness is much smaller than pitch) and when the pitch becomes comparable with the thickness (when some modes hybridize).

One challenge in our resonator design was matching the impedance over this range of frequencies. In the end, all resonators are matched to 200 Ω input impedance at the mid-band frequency and with an aperture of 10λ. This choice results in 192 fingers for the smallest pitch of 500 nm and 44 fingers for the largest pitch of 4 µm.

### A. Layout

All the photolithographic steps described in this paper are performed with a PAS 5500-350C DUV Stepper tool (ASML, Netherlands) equipped with a 248 nm KrF laser source and a demagnification factor of 4x Reticle to wafer feature size. Reticles are made of fused quartz with a 90 nm Chrome mask layer. Their size is 6 x 6 inches wide and 0.25 inch thick. Reticles can be purchased from commercial vendors or can be written, as done in this paper, in-house with a laser writer (Heidelberg VPG200 i-line). The reticle's maximum Image Field (IF) corresponds to a chip size of 2.2x2.2 cm2. However, Reticle Management (ReMa) blades can be used to selectively expose only a subset of the total field. This allows us to place the designs for all different lithographic steps on the same reticle.

Though in principle (see next section), we only require three etching steps to fabricate the resonators (bottom metal patterning, IDT patterning, and resonator sidewall patterning), closer attention must be paid when designing the IDT layer. Indeed, to attain the best results, the optimal dose is different for a small IDT pitch and a large IDT pitch. That is why we split the IDT designs into three: one containing the small pitches (from 500nm to 1µm), one containing the large pitches (1.5µm to 4.5µm), and a third containing the contact pads. Considering all these different layers, our reticle provides chip sizes of 17x3 mm. 83 chips are fit on the wafer, with a space in the wafer center to perform XRD measurements and optical endpoint detection during etching steps.

In each of those 83 chips, we include de-embedding open and short structures to evaluate parasitic resistance and capacitance impact on our measurements. Still, in the scope of this paper, all measurements shown are calibrated but not de-embedded.

### B. Alignment protocol

Our process requires a 3-layer stack in its simplest form, and the alignment needs to be below 100 nm for overlay accuracy. The typical alignment strategy for the used tool is based on Through-The-Lenses (TTL) alignment on Primary Marks (PM). However, this is not viable for our process since it is always necessary to align to clean marks on silicon around 135 nm deep. Instead, we investigate two alternatives: (i) 3D-Align[13], and (ii) Advanced Technology using High-order Enhancement of alignment (ATHENA). The former option allows for a more flexible stack choice on the front side since the alignment marks are patterned on the wafer's backside. This frees from the requirement of topside material choice but requires patterning of the wafer backside, adding complexity to the process. In addition, the periscope setup to align on the backside reduces the overlay accuracy. The latter option, ATHENA, relies on the 3rd, 5th, and 7th-order diffraction of two light sources for alignment [14], [15]. ATHENA uses specific segmented marks called Versatile Scribe line Primary Marks (VSPM) to maximize each diffraction order signal but does not require precise etch depth in silicon or backside processing. After testing, we achieve a <80 nm accuracy for each layer with the ATHENA system and <150nm for the 3D alignment. Thus, ATHENA is a better option for our process. Each photolithography step contains images for the devices, alignment marks, and Vernier test structures to measure the optical alignment of the overlay between different layers.

IDT is patterned to define the resonator wavelengths. Then, a SiO2 hard mask is deposited and patterned, and the AlScN is etched with IBE. The final step of the process is stripping the hard mask and releasing the devices.

### A. Bottom layer deposition and photolithography

The first step is the sputter deposition of the 25 nm Pt bottom metal electrode with a 10 nm adhesion-promoting layer (Fig. 2.a). This paper investigates Ti and AlN adhesion layers, with Ti deposited at 350 °C and AlN deposited at 300 °C, while the Pt electrode is always deposited at 350 °C. Deposition of the piezoelectric and metal stack is always performed with a Spider600 single-target, multi-chamber sputtering cluster. For the photolithography of the first layer, we follow the protocol introduced in[12]. We use a resist stack composed of JSR M108Y on top of a Brewer DS-K101 Developable (Wet) Bottom Antireflective Coating (BARC). The thickness of the BARC is 60 nm, while the thickness of the photoresist is 400 nm. The Wet BARC is removed during the photoresist development in TMA238WA with a minimal undercut (Fig. 2.b). The photoresist undergoes reflow with a 90 s bake at 170 °C to ensure no metal redeposition on the sidewalls originates during Ion beam etching. We etch the Pt layer in a Veeco350 Ion Beam Etcher (IBE) with two different etch depths according to the adhesion layer: when using AlN as adhesion, we only etch the Pt and land on the AlN, while when using Ti, we etch away the adhesion layer, to land on bare Si. (Fig. 2.c).

### B. Bottom layer resist strip and AlScN deposition

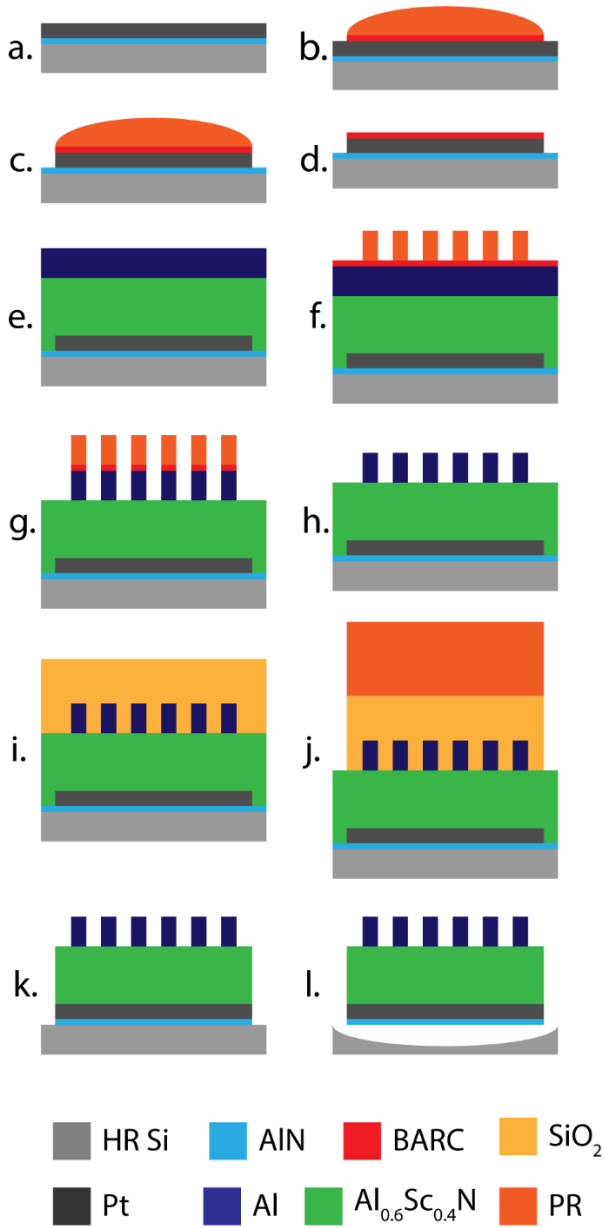

Fig. 2. Resonator fabrication process flow. a) deposition of adhesion layer and bottom Pt electrode; b) bottom layer DUV photolithography and reflow; c) IBE of bottom electrode; d) strip of burnt resist in plasma + 1165; e) stripping of BARC in O2 plasma and deposition of piezoelectric and top layers; f) DUV photolithography of top layer; g) BARC opening and electrode RIE; h) resist and BARC strip; i) deposition of SiO2 hard mask; j) DUV photolithography and etching of Hard mask k) IBE of AlScN layer and mask stripping; l) device release.

## III. FABRICATION PROCESS

The fabrication process for the resonators of this paper is illustrated in Fig. 2 and will be described in detail in this section.

All the resonators in this paper are deposited on Double Side Polished 100 mm (4-inch) High-resistive Silicon (HR-Si) wafers with <100> orientation and resistivity larger than $10\ k\Omega \cdot cm$. We first deposit the Pt bottom electrode and pattern it with photolithography and Ion Beam Etching; the resist is removed, and AlScN is sputter deposited on the metal electrodes, together with the top layer of Al. Again, using DUV photolithography, the top

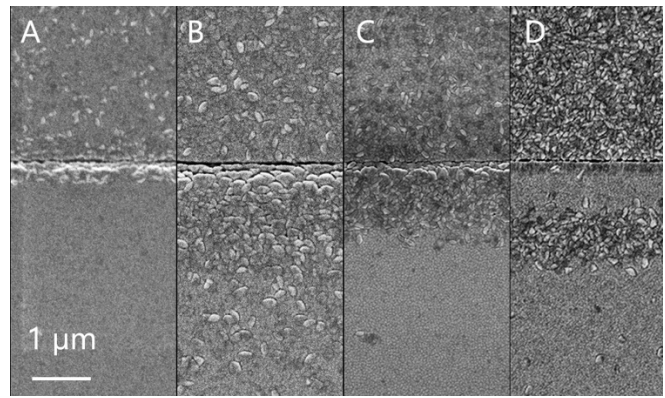

Fig. 3: Comparison of abnormal grain growth: A) ECI 3007 reference from[12]. B) Deposition on metal exposed with M108Y only, C) Deposition on metal exposed with DS-K101/M108Y, D) same as C) with an extra 30 s dip-HF clean step.

As mentioned in our previous work [12], removing photoresist post-etch residues is crucial in ensuring no particles remain on the metal surface, as they can cause Abnormally Oriented Grains (AOGs) on the AlScN thin film (Fig. 3). The fabrication process within this paper utilizes a different set of resists than in [12]. Indeed, when applying the same protocol optimized in [12] (1 min $O_2$ plasma / 10 min of Remover 1165 bath / 3 min of $O_2$ plasma), we observe that the first 1 min $O_2$ plasma cleaning step suffices to remove the resist entirely, therefore causing burnt residues to stick to the

surface(Fig. 3.b). Instead, we replace the first step in that protocol with a 1 min Forming Gas (N2/5%H2 mixture) in an ESI 3511 Asher on a heated plate at 120 °C. The measured ashing rate for this temperature is around 100 nm/min, making sure the 400 nm thin resist is not fully removed. In addition, we also add a layer of DS-K101 Bottom Antireflective Coating (BARC) under the M108Y resist. While M108Y is removed in the Remover 1165, BARC is unaffected, allowing a protection layer to cover the metal from any particle in the bath during the wet stripping phase (Fig. 2.d). BARC can then be removed in O$_2$ plasma, yielding a very clean metal surface that, in turn, produces a piezoelectric film with a lower number of AOG (Fig. 3.c). To explore if BARC residues might be present on the surface of the metal, an additional cleaning step could be introduced based on diluted HF (49%HF/H$_2$0 1:50) for 30 s. This results in a marginal reduction of AOG (Fig. 3.d). Importantly, since HF rapidly etches Ti, this latter test is performed on wafers that have AlN as an adhesion layer below Pt, which we have shown to be efficient for adhesion and Pt preparation [16]. After patterning and cleaning, we deposit the piezoelectric Al$_{0.6}$Sc$_{0.4}$N layer following the optimized conditions from [12], i.e., with a substrate temperature of 300 °C and gas flows of 10 sccm of Ar and 30 sccm of N$_2$. The thickness of the deposited layer is 400 ± 20 nm film thickness, with a maximum thickness of 420 ± 5 nm in the wafer center (variability is wafer-to-wafer, considering the same deposition conditions and time. XRD Rocking curve FWHM of the AlScN layers is 1.8° ± 0.15°. Without vacuum breaking, we deposit the top Al electrode at a temperature of 100 °C with a thickness ranging from 50 to 100 nm (**Fig. 2**.e). This results in Al resistivity of 32.4 Ω·nm. Note that the measurement of the AlScN layer is characterized with a Woolam RC2 ellipsometer and, since the aluminum top electrode absorbs incoming light, the measurement of the thickness is done after etching the top electrode and on dummy test wafers dedicated solely to thickness verification.

*C. Top electrode photolithography and etching*

As explained in the previous section, the finger dimensions in our designs range from 250 nm (for the devices with 1 µm wavelength and 500 nm pitch) to 2.25 µm (for the devices with 9 µm wavelength and 4.5 µm pitch). With this relatively large range, it is recommended to separate different devices into different layers (as explained in the previous section). The exposure can then be done seamlessly, taking advantage of the stepper's ability to align several exposures with the same image accurately. The three levels correspond to: (i) the contact pads, where the large size was not demanding in terms of exposure; (ii) the larger wavelength devices, with linewidth from 750 nm to 2.25 µm; and (iii) the smaller wavelength devices with linewidth from 250 to 500 nm. Critical dimensions measurements done with ProSEM software on dose tests show that the necessary dose to achieve a 0 deviation from fabrication to layout (in our dense array designs) is 20.50 mJ/cm$^2$ for all finger sizes except for the 250nm fingers that require 21.75 mJ/cm$^2$, as visible in Fig. 4.a.

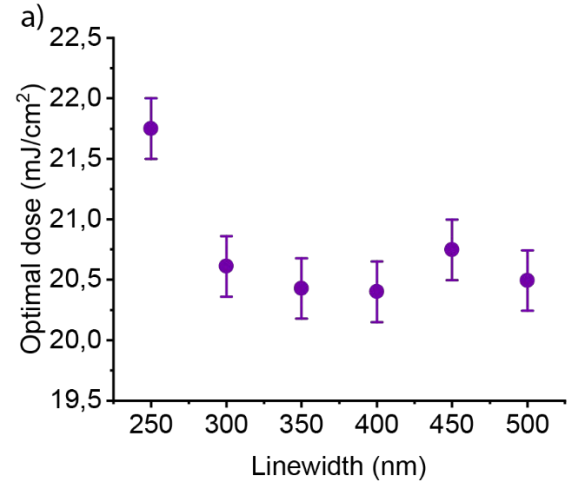

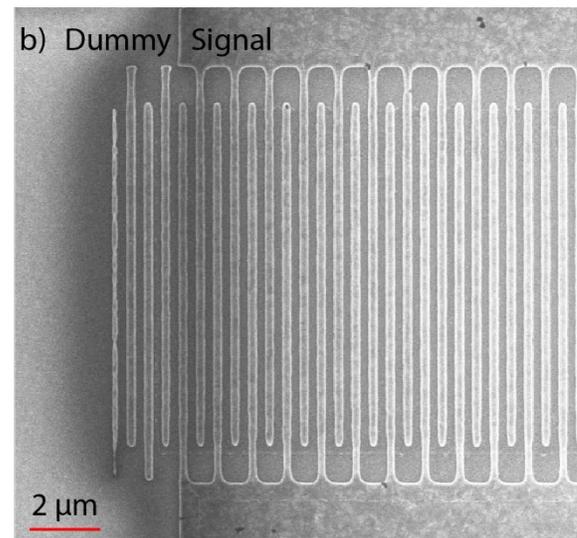

Fig. 4. a) Optimal dose to achieve 0 deviation from fabricated to layout fingers for the linewidth of the "small" devices, showing how the smaller linewidth requires a larger dose than the rest. B) SEM picture of fingers for a pitch of 500 nm, showing the uniform lithography of the signal fingers and the progressive overexposure of the dummies.

This means that the design splitting into different layers could be optimized with respect to our original choice (where we lumped 250-500 nm fingers in the same layer). Importantly, since the fingers form a dense array of structures, one needs to be careful when designing the electrode array's final region since the structures' density drops. To bypass this issue, in a similar way to what is currently done for transistor matching [17], extra fingers ("Dummy" fingers) are placed after the busbars are finished. They are not connected to the busbars and are only needed for the lithography of the electrodes. FIG. 4.b clearly shows how the electrode size changes when reaching the most external fingers, which are overexposed due to the reduction in finger density. Thanks to these dummy fingers, we can ensure a constant width throughout the busbar. Aluminum is the material chosen for our top electrode to minimize resistive and mechanical losses. However, with the use of Al, two points of concern arise. First, a BARC layer is required for repeatable lithography. Second, as it is widely known, aluminum is sensitive to the most common developers (e.g., TMA238WA), which will isotropically damage the electrode fingers during resist development, resulting in an irregular electrode shape (Fig. 5.a). The selected solution in this paper is to use a non-developable BARC, DUV42-P from Brewer Science. Such BARC is not opened during development, thus stopping the TMA238WA from etching the Al electrodes (Fig. 2.f) and it is instead later removed using a dry-etching process, yielding much straighter fingers (Fig. 5.b).

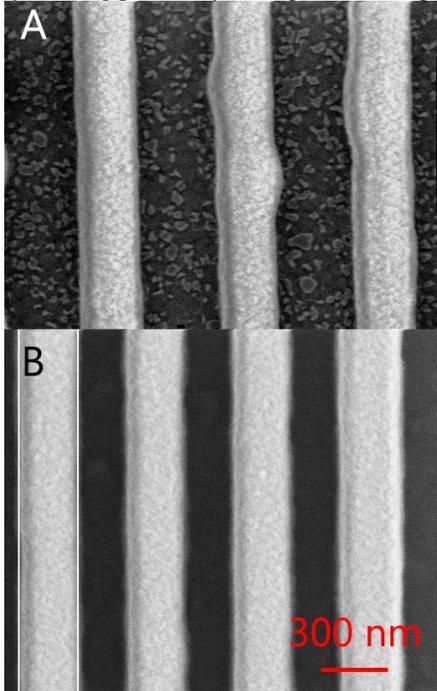

Fig. 5: Impact of TMA238WA on Al fingers: A) without BARC the finger shape is irregular, B) with BARC the fingers are more straight.

We open the BARC using a C4F8/O2 plasma (SPTS, UK). Subsequently, the electrodes are etched in Cl2/BCl3 plasma (Oxford, UK) using 150W of power applied to the substrate to achieve a high etch rate, finishing with a "soft landing" step using 50 W of bias. This, coupled with optical emission spectroscopy and laser interferometry end-point detection, allows us to land on AlScN without over-etching.
After Al etching, the wafers are immediately rinsed in DI water to reduce HCl corrosion[18], and then the remaining resist is stripped in Forming Gas to not oxidize the Al surface (Fig. 2.h).. The ashing rate at a plate temperature of 250 °C is 400 nm/min.

### D. AlScN patterning and device release

After patterning the top electrode, we sputter an 800 nm $SiO_2$ hard mask to provide a sufficient etching budget for the complete patterning of the AlScN layer (Fig. 2.i). Photolithography is done with an 1100 nm thick M35G photoresist. Due to the large feature sizes in this layer, no BARC is necessary in this step. The $SiO_2$ mask is opened in a $C_4F_8/H_2/He$ plasma (Fig. 2.j).
After opening the hard mask, AlScN etching is done in a Nexus350 Ion Beam Etcher (IBE, Veeco, USA) with a beam acceleration voltage of 500 V and a beam current of 800 mA/cm$^2$ using pure physical milling from Ar ions.

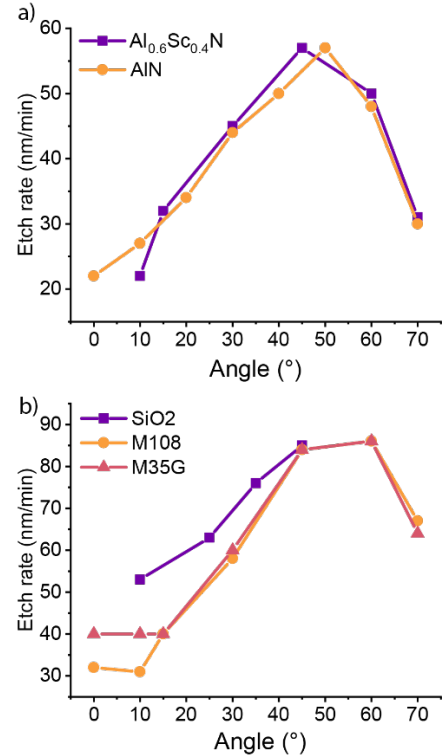

Fig. 6: Ion milling rates for a) Piezoelectric films, showing similar etching rate for AlN and $Al_{0.6}Sc_{0.4}N$ and b) Masks, showing that both photoresists used for the fabrication have a slower etch rate than $SiO_2$, potentially making a hard mask unnecessary..

The choice of IBE compared to $Cl_2$ ICP RIE for the AlScN layer etching comes from the flexibility of the process since the etch rate for $Al_{0.6}Sc_{0.4}N$ is very close to the etch rate for undoped AlN (Fig. 6.a). At the same time, when using RIE, it has been shown that the etch rate significantly decreases with the increase of Sc concentration [19], [20], [21]. Concerning the masking materials, the ion milling rate of the two photoresists is close, if not better, to the milling rate of the $SiO_2$ used as a hard mask (Fig. 6.b). The need for the hard mask comes from the thickness constraint of 1100 nm for the M35G photoresist. The hard mask can be avoided with a thicker photoresist or a thinner piezoelectric layer, and the fabrication process can be further streamlined. To etch the AlScN, we use a pulsed recipe composed of 3 steps: 1 minute at a 10 ° angle, 30 seconds at 45 °, and 30 seconds at 70 °, repeated 13 times to overetch into the Si substrate. This recipe results in a sidewall of 88 °, as visible in Fig. 7.a.

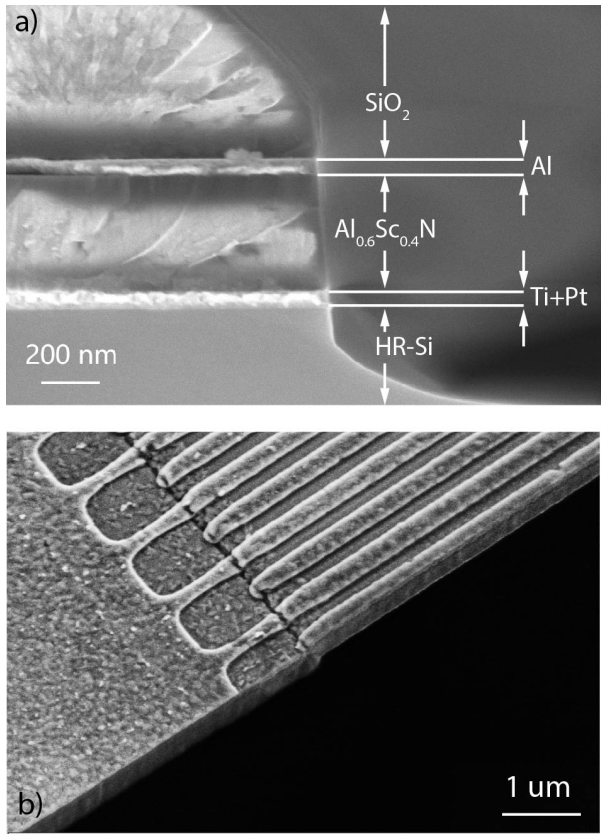

Fig. 7 a) Sidewall of AlScN after IBE, b) Detail of suspended resonator after XeF2 release.

After IBE, the remaining photoresist is stripped in $O_2$ plasma, and the hard mask is stripped in an anhydrous vapor HF gas SPTS uEtch etching tool (FIG. 2.k). Compared to wet stripping (Diluted HF, BHF, etc…), anhydrous HF does not attack the Al fingers underneath the hard mask nor the Ti under the bottom Pt (if present). The resonator is then released in $XeF_2$ gas etching, using 50 pulses of gas, each 45 seconds long, with 7.5 Torr of chamber pressure. A detail of the suspended resonator is visible in Fig. 7.b: the dummy fingers used during the photolithography of the IDT are etched away during the resonator patterning step

## IV. DEVICE CHARACTERIZATION

After release (Fig. 8.a), the devices are probed on an MPI150 probe station with GSG probes using an RS-ZNB20 VNA with a power of -10 dBm. Before measurement, a Short-Open-Load-Through (SOLT) calibration removes parasitics from the cables.

### A. Resonator electrical response

Measurements are then fitted using a multi-branch modified Butterworth-Van Dyke (mBVD) model (see Fig. 8.b for a comparison of measured resonator admittance and its fitted mBVD model) to extract the electric lumped elements and the performance parameters like quality factor and coupling for each resonance mode. To avoid confusion between main resonances and spurious modes, the measurements are compared with the FEM simulations to correctly address the wanted A0, S0, A1, and S1 modes. Since the devices' wavelengths range from 9 µm to 1 µm, the S0 mode can cover a frequency range from 700 MHz to 5 GHz, as visible in FIG. 9.

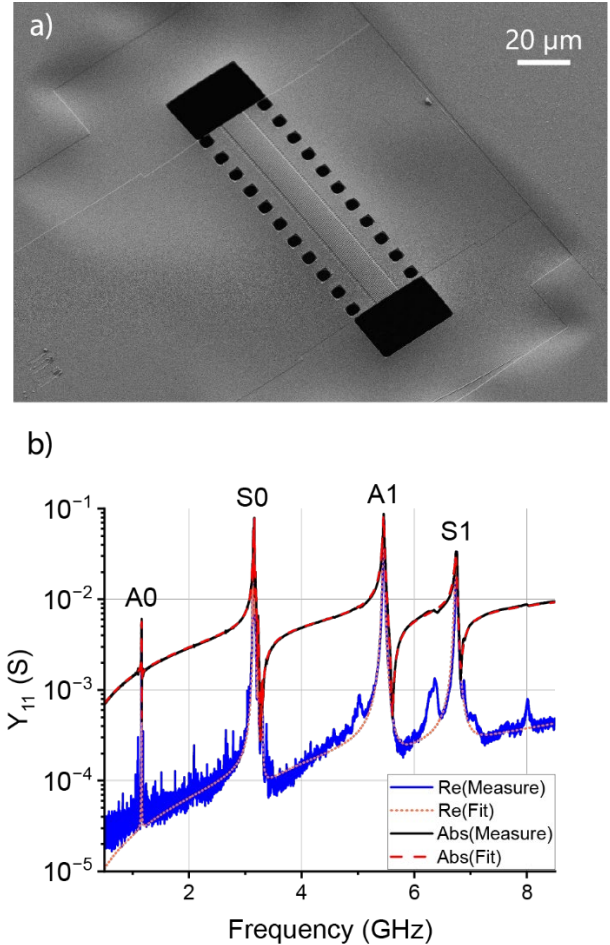

Fig. 8: a) SEM image of suspended resonator after release b) Comparison of measured resonator admittance and its fitted mBVD model. The resonator's pitch is 1 µm.

### B. Dispersion curves

To investigate the mode dispersion, we measure a batch of devices with the same aperture/wavelength ratio of 10 and gap width of lambda/2, with a different pitch, resulting in different wavenumber; our available electrode pitch range goes from 500 to 4500 nm. Results are visible in FIG. 10, showing the impact of wavelength on resonance frequency being more significant for A0 and S0 modes, as they are mostly pitch-defined, compared to the A1 and S1 modes, which are mostly thickness-defined.

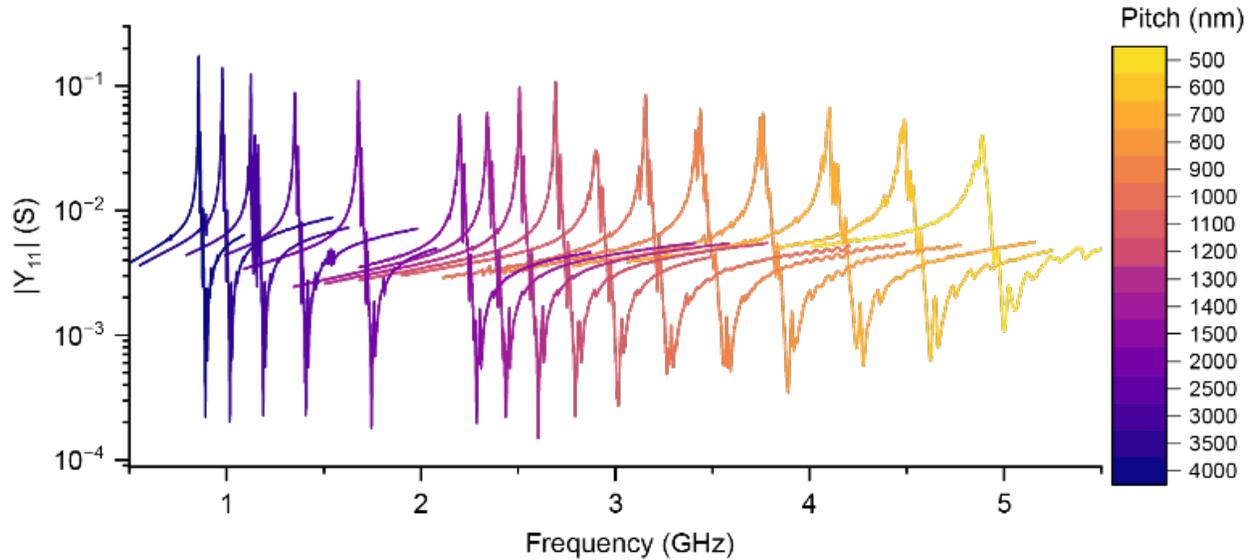

Fig. 9: Admittance of the S0 mode for the investigated electrode pitches.

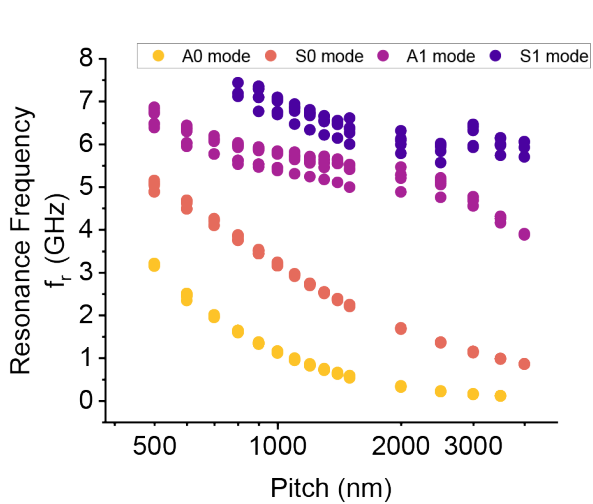

Fig. 10: Dispersion on the four primary modes across the wafer surface.

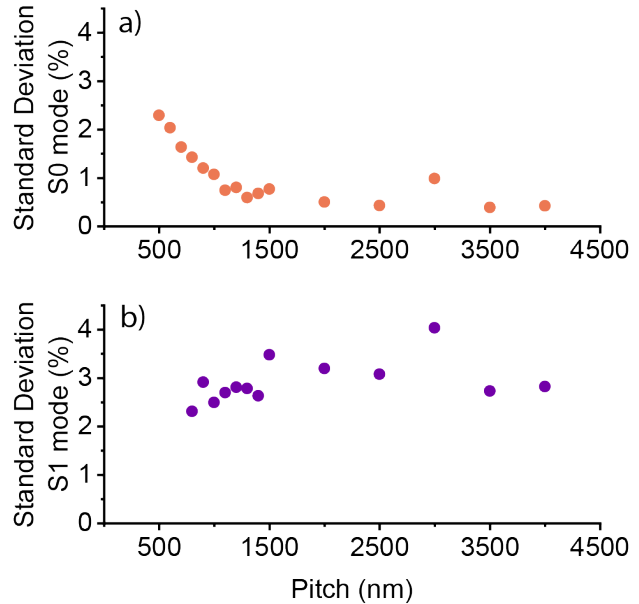

Fig. 11: Relative standard deviation of the resonance frequency for the a) S0 and b) S1 modes.

### C. Frequency deviation over wafer surface

Taking advantage of the high throughput of the DUV stepper, the device chip is repeated 93 times for each wafer. This allows us to investigate the deviation of resonance frequencies to quantify the reproducibility of the process. The results of this study are visible in Fig. 11. The S1 mode (Fig. 11.b), mainly being thickness defined, has a 3% relative standard deviation of resonance frequency across the whole pitch range, which is expected since the thickness has a similar standard deviation across the wafer surface. On the contrary, the S0 mode (Fig. 11.a), mostly lithography-defined, has a remarkable standard deviation of <1% for pitch values larger than 1 μm. Below 1 μm, the deviation increases due to an interaction of lateral and thickness modes [9], [22]. The accuracy of the resonators with smaller pitches can be improved by improving the deposition uniformity, for instance, by using ion trimming, as reported on LiNbO3 resonators[23].

### D. Evolution of Q and $K_{eff}^2$

By fitting the measured resonators over the wafer surface with the mBVD model, we can study the evolution of the quality factor and piezoelectric coupling coefficient against the resonance frequency for each mode. In this paper, we define the resonance (loaded) quality factor $Q_r = \frac{1}{\omega_r C_m (R_m + R_s)}$ and effective coupling coefficient $k_{eff}^2 = \frac{C_m}{C_m + C_0}$, following the IEEE definition [24]. Quality factor results are visible in Fig. 12.a: $Q_r$ of the A0 mode is noticeably higher than for the other modes and tends to increase with increasing resonance frequency, reaching a stable value of 700 for pitch above 1 μm, corresponding to a h/λ value of 0.2. The quality factors for the other modes follow a common trend, starting at around 300 for lower frequencies and lowering to 100 for higher frequencies. In the scope of this paper, we do not work to optimize the quality factor, and thus, it results in moderate values of $Q_r$. There are some strategies to improve $Q_r$ that we could implement by simply reducing $R_s$, or by working on the right metal combination for us to optimize the value of $R_m$. Results for $k_{eff}^2$ are shown in Fig. 12.b. In this case, the coupling of the A0 mode is low, not reaching 1% for all of the analyzed pitches, while for the S0 mode, it slowly decreases with frequency going from 8% to 6%. By contrast, S1 and A1 modes show a maximum coupling with a specific pitch value. As Lin et al. [25] demonstrated, the $k_{eff}^2$ of the S0 mode of a Lamb wave resonator with a floating metal bottom electrode decreases with a larger thickness-to-pitch ratio. Zou's thesis [26] shows that for AlN Lamb wave resonators with a floating metal bottom electrode, the coupling of S1 modes is larger than the one for S0 modes only for smaller h/λ values, coinciding in this paper's case with the larger pitches, where the S1 mode coupling surpasses the one of S0 for any pitch larger than 1 µm. A1 mode shows in this paper a lower coupling than S0 in almost all cases, but for a pitch of 1 µm, it attains a maximum value of 6% comparable to the one of S0.

## V. Conclusion

This paper shows a process flow to fabricate suspended Lamb wave resonators with a lithographic tool that could be used in mass production. The objective is to bridge the gap between fabrication methodologies oriented toward research and those more aligned with industrial applications' high throughput requirements. Our performance analysis shows that frequency deviation mostly depends on thickness non-uniformity rather than lithographic resolution limits and that this process issue can be solved by optimizing the thickness control of the piezoelectric layer. Streamlining the mask fabrication process by in-house writing helps in small-scale production and can be easily scaled to mass production, given that the DUV Stepper used is an industry-grade machine. The resonator performances can be significantly improved by using all the already existing design solutions used to limit spurious mode formation and reduction of parasitics, such as apodization or piston-mode operation[27], [28], [29]. It is worth noting that the employed photoresists can be written with electron-beam lithography[30], allowing for a mixed approach where research prototypes could be fabricated with the flexibility of e-beam[31]. Then, the final design can be quickly spun to mass production levels. Future developments of this process would benefit significantly from the previous 30 years of development of dry and immersion photolithography routinely employed in IC production, such as multi-patterning[32], to increase wafer resolution without using more advanced and design-restrictive photolithographic processes.

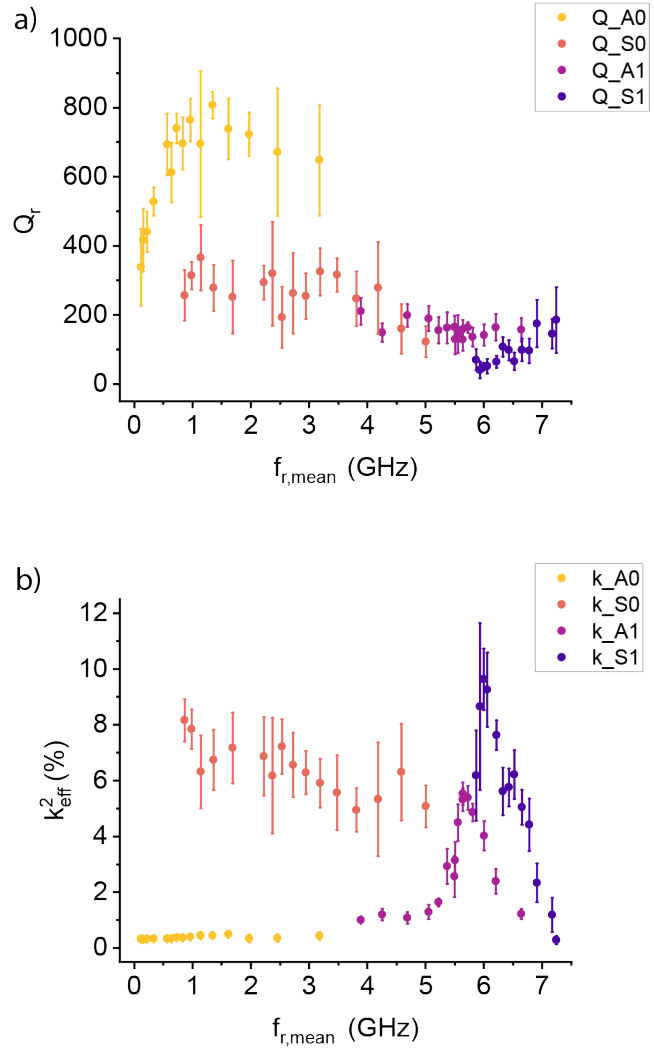

Fig. 12: average value and deviation across the wafer surface of a) Qr and b) $K_{eff}^2$ against the average resonance frequency for each mode.

## Acknowledgment

T The authors thank the staff of the EPFL Center of Microtechnology for their continuous assistance in process development and tool maintenance. A special acknowledgment goes to Niccolò Piacentini for all the invaluable help with DUV Lithography and mask setup and to Joffrey Pernollet for the help with etching and stripping tools recipes. The author M.L. thanks Yara Abdelaal for kindly providing the etch rate measurements for AlN.

## References

[1] "Status of the MEMS Industry 2023," Yole Group. Accessed: May 07, 2024. [Online]. Available: https://www.yolegroup.com/product/report/status-of-the-mems-industry-2023/

[2] A. Lozzi, M. Liffredo, E. T.-T. Yen, J. Segovia-Fernandez, and L. G. Villanueva, "Evidence of Smaller 1/F Noise in AlScN-Based Oscillators Compared to AlN-Based Oscillators," *J.*

*Microelectromechanical Syst.*, vol. 29, no. 3, pp. 306–312, Jun. 2020, doi: 10.1109/JMEMS.2020.2988354.

[3]　R. Ruby, "A decade of FBAR success and what is needed for another successful decade," in *2011 Symposium on Piezoelectricity, Acoustic Waves and Device Applications (SPAWDA)*, Dec. 2011, pp. 365–369. doi: 10.1109/SPAWDA.2011.6167265.

[4]　J. Wang, M. Park, S. Mertin, T. Pensala, F. Ayazi, and A. Ansari, "A Film Bulk Acoustic Resonator Based on Ferroelectric Aluminum Scandium Nitride Films," *J. Microelectromechanical Syst.*, vol. 29, no. 5, pp. 741–747, Oct. 2020, doi: 10.1109/JMEMS.2020.3014584.

[5]　Z. Schaffer, P. Simeoni, and G. Piazza, "33 GHz Overmoded Bulk Acoustic Resonator," *IEEE Microw. Wirel. Compon. Lett.*, vol. 32, no. 6, pp. 656–659, Jun. 2022, doi: 10.1109/LMWC.2022.3166682.

[6]　R. Vetury et al., "A Manufacturable AlScN Periodically Polarized Piezoelectric Film Bulk Acoustic Wave Resonator (AlScN P3F BAW) Operating in Overtone Mode at X and Ku Band," in *2023 IEEE/MTT-S International Microwave Symposium - IMS 2023*, Jun. 2023, pp. 891–894. doi: 10.1109/IMS37964.2023.10188141.

[7]　S. Yandrapalli, S. E. K. Eroglu, V. Plessky, H. B. Atakan, and L. G. Villanueva, "Study of Thin Film LiNbO3 Laterally Excited Bulk Acoustic Resonators," *J. Microelectromechanical Syst.*, vol. 31, no. 2, pp. 217–225, Apr. 2022, doi: 10.1109/JMEMS.2022.3143354.

[8]　G. Giribaldi, L. Colombo, P. Simeoni, and M. Rinaldi, "Compact and wideband nanoacoustic pass-band filters for future 5G and 6G cellular radios," *Nat. Commun.*, vol. 15, no. 1, p. 304, Jan. 2024, doi: 10.1038/s41467-023-44038-9.

[9]　C. Cassella, Y. Hui, Z. Qian, G. Hummel, and M. Rinaldi, "Aluminum Nitride Cross-Sectional Lamé Mode Resonators," *J. Microelectromechanical Syst.*, vol. 25, no. 2, pp. 275–285, Apr. 2016, doi: 10.1109/JMEMS.2015.2512379.

[10]　G. Q. Zhang and A. Roosmalen, Eds., *More than Moore*. Boston, MA: Springer US, 2009. doi: 10.1007/978-0-387-75593-9.

[11]　"What's driving the next renaissance in dry lithography?," ASML. Accessed: May 07, 2024. [Online]. Available: https://www.asml.com/en/news/stories/2019/driving-the-next-renaissance-in-dry-lithography

[12]　M. Liffredo, N. Xu, S. Stettler, F. Peretti, and L. G. Villanueva, "Piezoelectric and elastic properties of Al0.60Sc0.40N thin films deposited on patterned metal electrodes," *J. Vac. Sci. Technol. A*, vol. 42, no. 4, p. 043404, May 2024, doi: 10.1116/6.0003497.

[13]　H. W. van Zeijl and P. M. Sarro, "Alignment and overlay characterization for 3D integration and advanced packaging," in *2009 11th Electronics Packaging Technology Conference*, Dec. 2009, pp. 447–451. doi: 10.1109/EPTC.2009.5416504.

[14]　C.-B. Tan, S.-H. Yeo, H. P. Koh, C. K. Koo, Y. M. Foong, and Y. K. Siew, "Evaluation of alignment marks using ASML ATHENA alignment system in 90-nm BEOL process," in *Metrology, Inspection, and Process Control for Microlithography XVII*, SPIE, Jun. 2003, pp. 1211–1218. doi: 10.1117/12.487734.

[15]　G. M. Pugh and M. R. Giorgi, "Evaluation of ASML ATHENA alignment system on Intel front-end processes," in *Metrology, Inspection, and Process Control for Microlithography XVI*, SPIE, Jul. 2002, pp. 286–294. doi: 10.1117/12.473468.

[16]　K. M. Howell, W. Bashir, A. De Pastina, R. Matloub, P. Muralt, and L. G. Villanueva, "Effect of AlN seed layer on crystallographic characterization of piezoelectric AlN," *J. Vac. Sci. Technol. A*, vol. 37, no. 2, p. 021504, Mar. 2019, doi: 10.1116/1.5082888.

[17]　T. Datta and P. Abshire, "Mismatch compensation of CMOS current mirrors using floating-gate transistors," in *2009 IEEE International Symposium on Circuits and Systems*, May 2009, pp. 1823–1826. doi: 10.1109/ISCAS.2009.5118132.

[18]　K. Hirose, H. Shimada, S. Shimomura, M. Onodera, and T. Ohmi, "Ion-Implanted Photoresist and Damage-Free Stripping," *J. Electrochem. Soc.*, vol. 141, no. 1, p. 192, Jan. 1994, doi: 10.1149/1.2054683.

[19]　Z. Tang, G. Esteves, J. Zheng, and R. H. Olsson, "Vertical and Lateral Etch Survey of Ferroelectric AlN/Al1−xScxN in Aqueous KOH Solutions," *Micromachines*, vol. 13, no. 7, p. 1066, Jul. 2022, doi: 10.3390/mi13071066.

[20]　R. M. R. Pinto, V. Gund, C. Calaza, K. K. Nagaraja, and K. B. Vinayakumar, "Piezoelectric aluminum nitride thin-films: A review of wet and dry etching techniques," *Microelectron. Eng.*, vol. 257, p. 111753, Mar. 2022, doi: 10.1016/j.mee.2022.111753.

[21]　A. Lozzi, E. Ting-Ta Yen, P. Muralt, and L. G. Villanueva, "Al0.83Sc0.17N Contour-Mode Resonators With Electromechanical Coupling in Excess of 4.5%," *IEEE Trans. Ultrason. Ferroelectr. Freq. Control*, vol. 66, no. 1, pp. 146–153, Jan. 2019, doi: 10.1109/TUFFC.2018.2882073.

[22]　O. Kaya, X. Zhao, and C. Cassella, "An Aluminum Scandium Nitride (AL0.64SC0.36N) Two-Dimensional-Resonant-Rods Delay Line with 7.5% Bandwidth and 1.8 DB Loss," p. 4.

[23]　V. Chulukhadze et al., "Frequency Scaling Millimeter Wave Acoustic Resonators using Ion Beam Trimmed Lithium Niobate," in *2023 Joint Conference of the European Frequency and Time Forum and IEEE International Frequency Control Symposium (EFTF/IFCS)*, Toyama, Japan: IEEE, May 2023, pp. 1–4. doi: 10.1109/EFTF/IFCS57587.2023.10272038.

[24]　R. Lu, M.-H. Li, Y. Yang, T. Manzaneque, and S. Gong, "Accurate Extraction of Large Electromechanical Coupling in Piezoelectric MEMS Resonators," *J. Microelectromechanical Syst.*, vol. 28, no. 2, pp. 209–218, Apr. 2019, doi: 10.1109/JMEMS.2019.2892708.

[25]　C.-M. Lin, V. Yantchev, J. Zou, Y.-Y. Chen, and A. P. Pisano, "Micromachined One-Port Aluminum Nitride Lamb Wave Resonators Utilizing the Lowest-Order Symmetric Mode," *J. Microelectromechanical Syst.*, vol. 23, no. 1, pp. 78–91, Feb. 2014, doi: 10.1109/JMEMS.2013.2290793.

[26]　J. Zou, "High-Performance Aluminum Nitride Lamb Wave Resonators for RF Front-End Technology," UC Berkeley, 2015. Accessed: May 03, 2024. [Online]. Available: https://escholarship.org/uc/item/9v995545

[27]　S. Stettler and L. G. Villanueva, "Transversal Spurious Mode Suppression in Ultra-Large-Coupling SH0 Acoustic Resonators on YX36°-Cut Lithium Niobate," *J. Microelectromechanical Syst.*, vol. 32, no. 3, pp. 279–289, Jun. 2023, doi: 10.1109/JMEMS.2023.3262021.

[28]　M. Solal, O. Holmgren, and K. Kokkonen, "Design, simulation, and visualization of R-SPUDT devices with transverse mode suppression," *IEEE Trans. Ultrason. Ferroelectr. Freq. Control*, vol. 57, no. 2, pp. 412–420, Feb. 2010, doi: 10.1109/TUFFC.2010.1421.

[29]　J. Zou, J. Liu, and G. Tang, "Transverse Spurious Mode Compensation for AlN Lamb Wave Resonators," *IEEE Access*, vol. 7, pp. 67059–67067, 2019, doi: 10.1109/ACCESS.2019.2908340.

[30]　D. Maillard, Z. Benes, N. Piacentini, and L. G. Villanueva, "Electron-beam lithography on M108Y and M35G chemically amplified DUV photoresists," *Micro Nano Eng.*, vol. 13, p. 100095, Nov. 2021, doi: 10.1016/j.mne.2021.100095.

[31]　F. Laulagnet et al., "Electron beam direct write lithography: the versatile ally of optical lithography," *J. MicroNanopatterning Mater. Metrol.*, vol. 22, no. 4, p. 041404, Jun. 2023, doi: 10.1117/1.JMM.22.4.041404.

[32]　Y.-K. Choi, T.-J. King, and C. Hu, "A spacer patterning technology for nanoscale CMOS," *IEEE Trans. Electron Devices*, vol. 49, no. 3, pp. 436–441, Mar. 2002, doi: 10.1109/16.987114.